\documentstyle[12pt,epsf]{article}
\date{}

\begin{document}

\title{\bf Static skyrmions in (2+1) dimensions}
\author{R. J. Cova\thanks{Permanent Address: Dept. de F\'{\i}sica
 FEC, Universidad del Zulia, Apartado 526, Maracaibo, Venezuela.}
 \and W. J. Zakrzewski \\
{\em Department of Mathematical Sciences, University of Durham}, \\
{\em Durham DH1 3LE, UK}}

\maketitle
\begin{abstract}
 In  the spirit  of previous papers, but using more general
 field configurations, the  non-linear $O(3)$ model in (2+1)-D,
 modified by the addition of both a potential-like term and
a Skyrme-like term, is considered. The instanton solutions
 are  numerically evolved in time
and some of their stability  properties studied. They are
 found to be stable, and a repulsive force is seen to exist
 among them.
These results, which are restricted to the case of zero-speed
systems, confirm those obtained in previous  investigations,
in which a similar problem was studied for a  different choice
of the potential-like term.
\end{abstract}

\section{Introduction}

In the past few years, $\sigma$-models in low dimensions have
 become
an increasingly important area of research, often arising as
approximate models in the contexts of both particle and
solid state
physics. They have been used in the construction of
high-$T_{c}$ superconductivity and the quantum Hall effect;
 in two Euclidean dimensions, they appear to be the
low-dimensional analogues of
four-dimensional Yang-Mills theories. But only very special
$\sigma$-models in (2+1)-D are integrable \cite{ward}, and the
physically relevant Lorentz-invariant models are not amongst them;
 in these cases,
recourse to numerical evolution must be made. The simplest
 Lorentz-invariant model in (2+1)-D is the $O(3)$ model, which
involves  three real scalar fields,
{\boldmath $\phi$}($x^{\mu}$)$\equiv$\{$\phi_{a}$($x^{\mu}$),
 $a$=1,2,3\},
with the constraint that {\boldmath $\phi$} lies on the unit
sphere $S^{(\phi)}_{2}$:
\begin{equation}
\mbox{\boldmath $\phi.\phi$}=1.
\label{c}
\end{equation}
 Subject to this constraint, the Lagrangian density and the
corresponding equations of motion are
 \begin{equation}
 {\cal L}_{\sigma}=\frac{1}{4}
(\partial_{\mu}\mbox{\boldmath $\phi$}).(\partial^{\mu}
\mbox{\boldmath $\phi$}),
\label{lp}
 \end{equation}
 \begin{equation}
\partial^{2}_{t}\mbox{\boldmath $\phi$}=[
-(\partial_{t}\mbox{\boldmath
$\phi$})^{2}+(\partial_{x}\mbox{\boldmath $\phi$})^{2}
+(\partial_{y}\mbox{\boldmath
$\phi$})^{2}]\mbox{\boldmath $\phi$}+\partial^{2}_{x}
\mbox{\boldmath
$\phi$}+\partial^{2}_{y}\mbox{\boldmath $\phi$}.
  \label{eqmp}
\end{equation}
 Note that we are concerned with the model in (2+1)-D:
 \begin{math} x^{\mu} \equiv (x^{0},x^{1},x^{2})=(t,x,y)
\end{math}, with the speed of light set equal to unity.

An alternative and convenient formulation of the  model
 is in terms of one independent complex field, $W$, related
to the fields $\phi_{a}$ via
\begin{equation}
W=\frac{1-\phi_{3}}{\phi_{1}+i\phi_{2}}.
\label{rpw}
\end{equation}
 In this formulation, the Lagrangian density and the
 corresponding equations of motion read (asterisk denotes
complex conjugation)
\begin{equation}
{\cal L}_{\sigma}=
\frac{\partial_{\mu}W\partial^{\mu}W^{*}}{(1+|W|^{2})^{2}},
\label{ldw}
\end{equation}
and
\begin{equation}
\partial^{2}_{t}W=
\partial^{2}_{x}W+\partial^{2}_{y}W
+\frac{2W^{*}[(\partial_{t}W)^{2}-(\partial_{x}W)^{2}
-(\partial_{y}W)^{2}]}{1+|W|^{2}}.
\label{eqmw}
\end{equation}
 The problem is completely specified by giving the boundary
conditions. As usual we take
\begin{equation}
\lim_{r \rightarrow \infty}
\mbox{\boldmath $\phi$}(r,\theta,t)
=\mbox{\boldmath $\phi$}_{0}(t),
\label{b}
\end{equation}
where {\boldmath $\phi$}$_{0}$($t$) is independent of the
polar angle $\theta$. In (2+1)-D this condition ensures a
finite potential energy, whereas in two Euclidean dimensions,
 {\em i.e.\/}, when {\boldmath $\phi$} is independent of time, this
 leads to the finiteness of the action, which is precisely the
 requirement for quantization in terms of path integrals.
 As shown by several people \cite{belavin,woo}, any rational function
 $W(z)$ or $W(z*)$, where $z=x+iy$, is a static solution of
 eq. (\ref {eqmw}). These are the instantons of the model,
 and can be regarded as {\em static} solitons of the same
 model in (2+1)-D. The simplest one-soliton solution,
 $W=\lambda z$ ($\lambda$ is a free parameter determining
the size of $W$), has been numerically  studied by Leese
{\em et.al\/} \cite{leese1}. When viewed as an evolving structure in
 (2+1)-D,  the soliton has been found to be unstable; any small
 perturbation, either explicit or introduced by the
discretization procedure, changes its size. This
instability is associated with the conformal invariance of
 the $O(3)$ Lagrangian in two dimensions.

The $O(3)$ solitons, however, can be stabilized through  a judicious
introduction of a scale into the model, thereby breaking its
 conformal invariance. As explained below, in the present
 article we study this problem following the methods of
 a previous investigation \cite{leese2}, but considering a different
 potential-like term and restricting ourselves to the  systems
initially at rest.
\pagebreak

\section{Skyrme model in (2+1) dimensions}

Using the $W$-formulation, our Skyrme model is defined by
 the Lorentz-invariant Lagragian density
\begin{eqnarray}
{\cal L}&=&{\cal L}_{\sigma}\nonumber \\
&-&2\theta_{1}[\frac{(\partial_{t}W^{*}
\partial_{y}W-\partial_{t}W\partial_{y}W^{*})^{2}
+(\partial_{t}W^{*}\partial_{x}W-\partial_{t}
W\partial_{x}W^{*})^{2}}{(1+|W|^{2})^{4}}\nonumber \\
&-&\frac{(\partial_{x}W^{*}\partial_{y}W
-\partial_{x}W\partial_{y}W^{*})^{2}}
{(1+|W|^{2})^{4}}]\nonumber \\
&-&4\theta_{2}\frac{|W-\lambda|^{8}}{(1+|W|^{2})^{4}},
\label{lds}
\end{eqnarray}
where ${\cal L}_{\sigma}$ is given by eq. (\ref {ldw})
and $\theta_{1}$,$\theta_{2}$ are real parameters  with
dimensions of length squared and inverse length squared,
 respectively; they introduce a scale into the model,
 which is no longer conformal invariant. If the size of
the solitons is appropriately chosen, it is energetically
unfavourable for the solitons to change it.
The $\theta_{1}$-term is the (2+1)-D analogue of the
Skyrme term, whereas  the $\theta_{2}$-term is a
potential-like one. Unlike the former, the latter term
 is highly nonunique \cite{azca}.

 The field equation corresponding to the above Lagrangian
can be cast into the form

\begin{eqnarray}
W_{tt}&=&
W_{xx}+W_{yy}
+\frac{2W^{*}[(W_{t})^{2}-(W_{x})^{2}-(W_{y})^{2}]}{1+|W|^{2}}
\nonumber \\
&-&\frac{4\theta_{1}}{(1+|W|^{2})^{2}}[2W^{*}_{tx}
W_{t}W_{x}+2W^{*}_{ty}W_{t}W_{y}-2W^{*}_{xy}W_{x}W_{y}
\nonumber \\
&+&W^{*}_{xx}(W_{y}^{2}-W_{t}^{2})+W^{*}_{yy}(W_{x}^{2}
-W_{t}^{2})-W^{*}_{tt}(W_{x}^{2}+W_{y}^{2})\nonumber \\
&+&W_{xx}(|W_{t}|^{2}-|W_{y}|^{2})+W_{yy}(|W_{t}|^{2}
-|W_{x}|^{2})+W_{tt}(|W_{x}|^{2}+|W_{y}|^{2})\nonumber \\
&+&W_{xy}(W_{x}^{*}W_{y}+W_{y}^{*}W_{x})-W_{tx}(W_{t}^{*}W_{x}
+W_{x}^{*}W_{t})-W_{ty}(W_{t}^{*}W_{y}
+W_{y}^{*}W_{t})\nonumber \\
&+&\frac{2W}{1+|W|^{2}}((W^{*}_{t}W_{y}-W^{*}_{y}W_{t})^{2}
+(W^{*}_{x}W_{t}-W^{*}_{t}W_{x})^{2}-(W^{*}_{x}W_{y}
-W^{*}_{y}W_{x})^{2})]\nonumber \\
&+&\frac{16 \theta_{2}|W-\lambda|^{2}}{(1+|W|^{2})^{3}},
\label{eqms}
\end{eqnarray}
where the notation
 $W_{x}\equiv \partial_{x}W$, $W_{xx}\equiv
\partial_{x}^{2}W$, {\em etc}., has been used.

It is straightforward to check that
\begin{equation}
W=\lambda \frac{z-a}{z-b}
\label{s}
\end{equation}
is a static solution of eq. (\ref {eqms}), provided the
following relation holds:
\begin{equation}
\lambda=\frac{\sqrt[4]{2 \theta_{1}/\theta_{2}}}{a-b}.
\label{fix}
\end{equation}

Now $\lambda$, which characterizes the size of the
instanton $W$, is no longer a free parameter. It is fixed
 by eq. (\ref{fix}). The instanton, with its size thus fixed,
is usually referred to as a `skyrmion'. Unlike the ordinary
 $O(3)$ case \cite{leese1}, where the instanton changes its size
as time elapses, we expect our skyrmion to be stable.
 In reference 5, a similar question was studied
 using $-4\theta_{2}/(1+|W|^{2})^{4}$ for
the potential-like
 term, with the Skyrme field given by $\lambda z$, where
 \begin{math} \lambda=\sqrt[4]{2 \theta_{1}/\theta_{2}}
\end{math}.

\section{Numerical setup and results}

To perform our numerical simulations we employ the fouth-order
 Runge-Kutta\linebreak method, and approximate the spatial
derivatives by finite differences; the Laplacian is evaluated
using the nine-point formula. All computations were performed
on the workstations at Durham, on a fixed 201$\times$201 lattice
 with spatial and time steps $\delta x$=$\delta y$=0.02
 and $\delta t$=0.005. Every few iterations we rescaled the
 fields \begin{math} \mbox{\boldmath $\phi$}\rightarrow
 \mbox{\boldmath $\phi$}/\sqrt{\mbox{\boldmath
$\phi.\phi$\,}\,}\, \end{math}.
 We also included along the boundary a narrow strip to absorb
the various radiation waves, thus reducing their  effects on
 the skyrmions via the reflections from the boundary. As time
elapses, this absorption of radiation manifests itself
 through a small decrease of the total energy, which
gradually stabilizes
as the radiation waves are gradually absorbed.

We choose the parameters to have the values:
 $\theta_{1}$=0.015006250, $\theta_{2}$=0.1250, $a$=0.75
and $b$=0.05 which, according to eq. (\ref {fix}),
 set $\lambda$=1.

\subsection{Results for one skyrmion}

The numerical evolution shows that the skyrmion's shape,
 of which a representative picture is given in the upper
 half of figure 1, practically does not change as a function
of time. The maximum of the total energy density, which is
 related to the soliton size, starts off at the value
 128.5 and, after some radiation waves are emitted,
 stabilizes at $\approx$ 129.5 (see lower half of figure 1).
 This is in agreement
with the analytical result, as the expression for the maximum
of the total energy density, namely,
 $E_{max}=\varepsilon(1+\theta_{1}\varepsilon)$,
 $\varepsilon=8(|\lambda|^{2}+1)^{2}/|\lambda(a-b)|^{2}$,
 yields 129.3, the `canonical size'.

 In figure 2 are exhibited some kinetic energy density waves.
They propagate out to the boundary at the speed of light,
 leaving the region where the skyrmion is located essentially
 free of kinetic energy. The smallness of the kinetic  energy
indicates that the field given by eq. (\ref {s}) remains
 almost perfectly static. For example, the amplitude of the
 total energy density, at $t$=0, is positioned at
 $z_{max}$=(0, 0.4000) and, for $t$=10, it has been slowly
shifted to (0, 0.4013).

 So, as expected, our skyrmion appears to be stable under
 perturbations brought about by the discretization procedure.
 These results agree with those of reference \cite{leese2}, where
the skyrmion was positioned at the centre of the grid, {\em i.e.},
 $z_{max}$=0, which, as one can easily check, corresponds to
 a skyrmion given by $W=\lambda z$ or $W=1/\lambda z$.

 In order to study the stability problem further, we are
 currently performing simulations with the
 same $\theta_{i}$-values
 but setting $\lambda \neq 1$. This corresponds to initial
 conditions that are no longer those of the static solution.
We hope some concrete results will be available in the future.

\subsection{Results for two skyrmions}

Let us now consider the case where two instantons are put
 on a lattice. We consider the field
\begin{equation}
W=\lambda \frac{z-a}{z-b}\frac{z+c}{z+d},
\end{equation}
with $\lambda$=1, $a=c=0.75$, $b=d=0.05$ and let it evolve.
 One can readily verify that the above field configuration is
 not a solution of the equations of motion, and hence it
 evolves when started off at rest.

 Our simulations  have shown that, at the initial
time, $E_{max}$ is well above the canonical size as
 determined in the last sub-section. As soon as the evolution
in time begins the solitons shake off some kinetic energy and
alter their size by getting broader;  $E_{max}$
decreases and then undergoes
 damped-oscillations around a value very close to the
 canonical one. As this happens,  the solitons slowly move
away from each other, exhibiting the presence of a repulsive
force between them.  By $t \approx 8$, the amplitude of the
above-mentioned oscillations is quite small, and $E_{max}$
eventually stabilizes near the expected canonical value.
Typical pictures for this case are displayed in figures 3 and 4.

\begin{figure}
\epsfverbosetrue
\centerline{\epsfbox{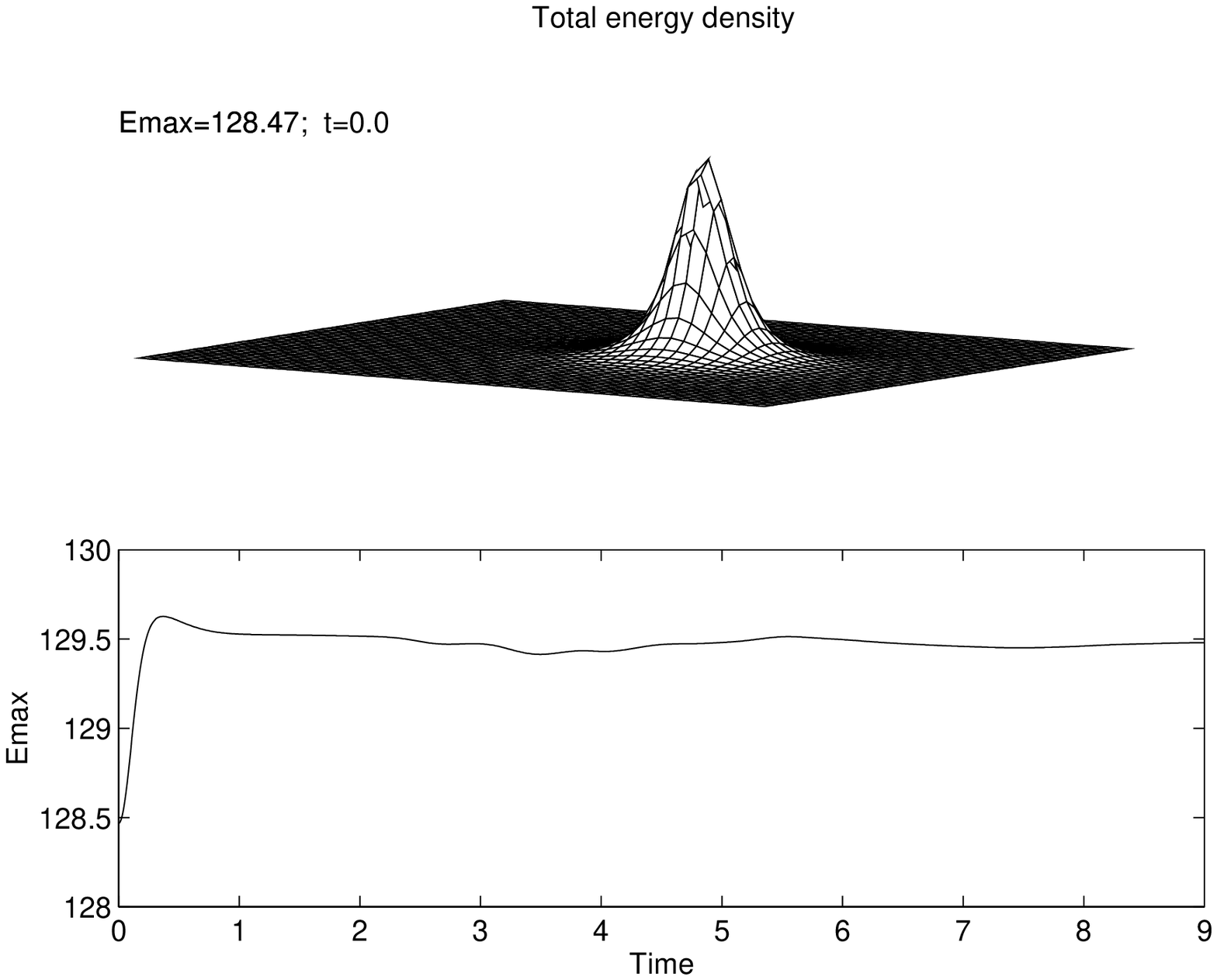}}
\caption{Total energy density for one skyrmion at the initial
time and the evolution of its peak.}
\end{figure}

\begin{figure}
\epsfverbosetrue
\centerline{\epsfbox{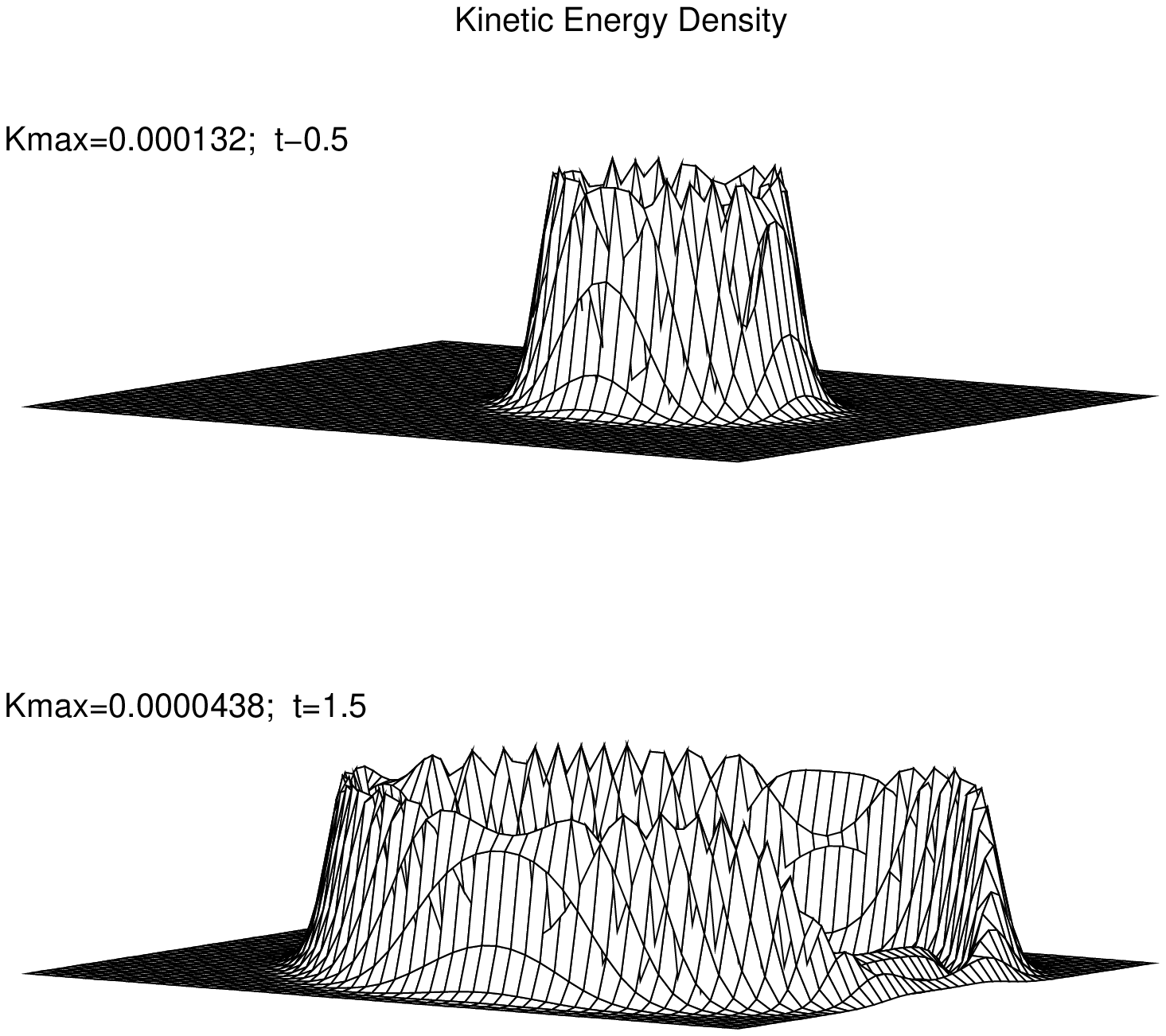}}
\caption{Kinetic waves travelling towards the boundary.}
\end{figure}

\begin{figure}
\epsfverbosetrue
\centerline{\epsfbox{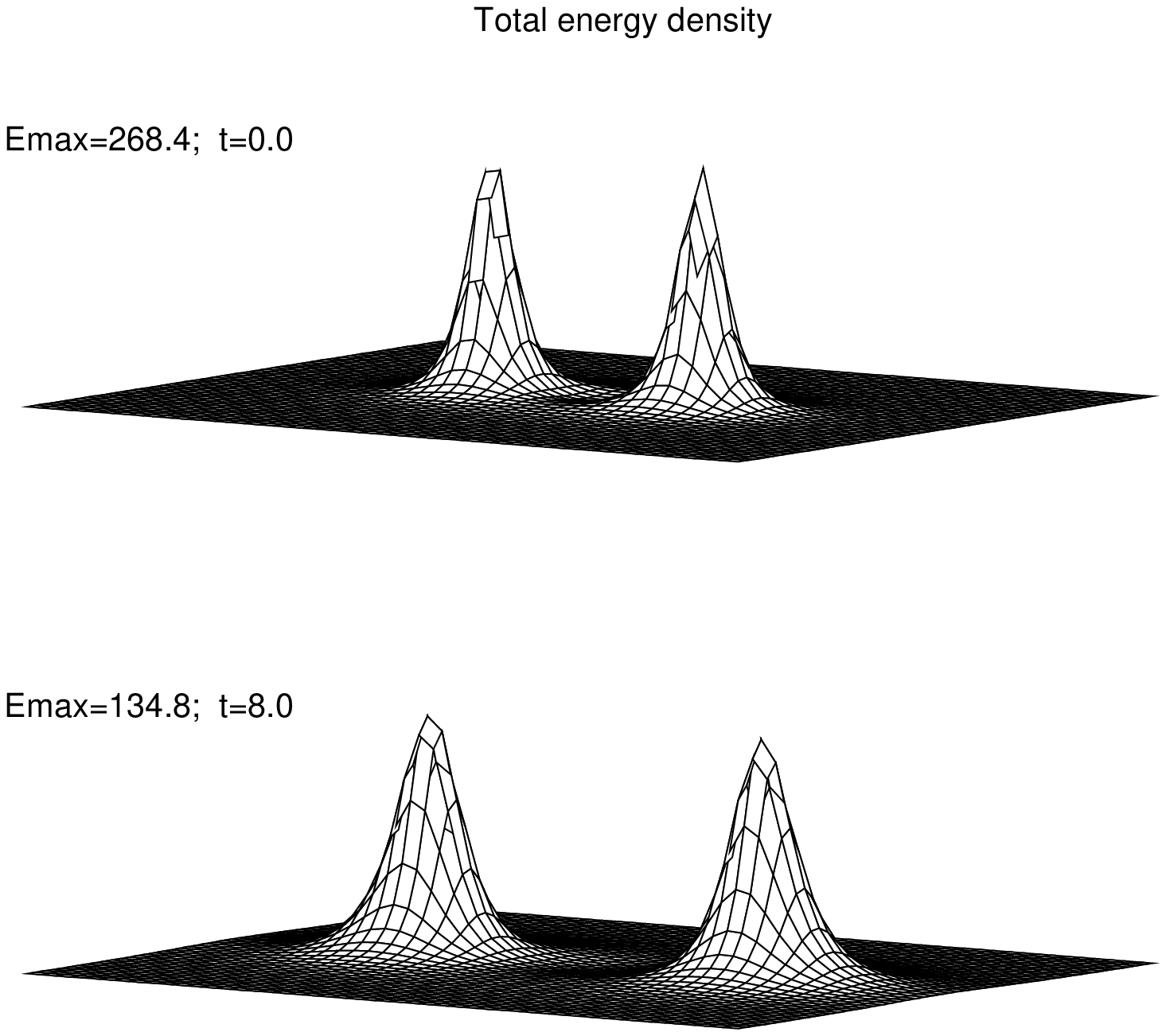}}
\caption{Two skyrmions. A repulsive force exists between them.}
\end{figure}

\begin{figure}
\epsfverbosetrue
\centerline{\epsfbox{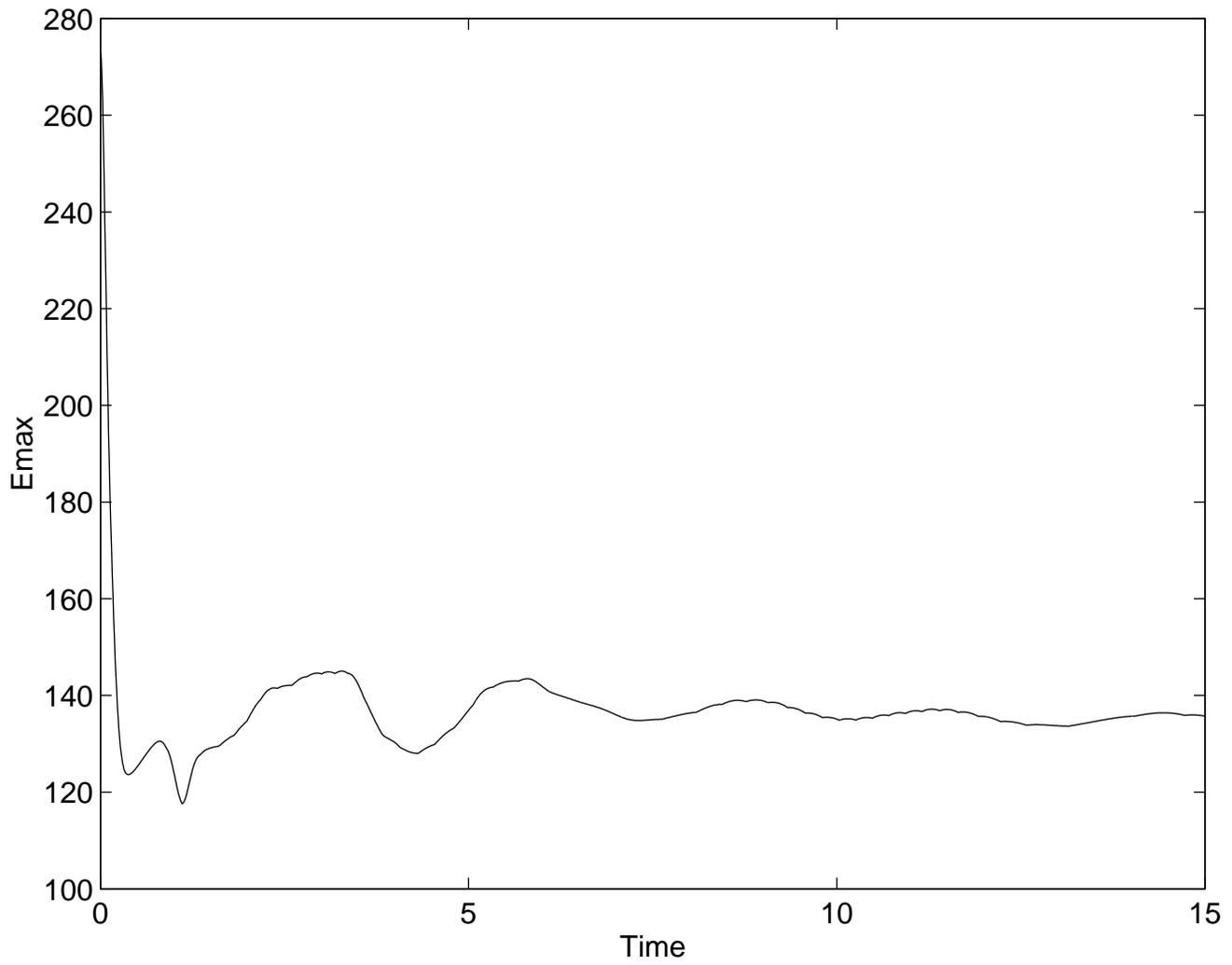}}
\caption{Evolution of the maximum of the total energy density
for the two-skyrmions case.}
\end{figure}

\newpage
\section{Conclusions}
We have performed a numerical study of the
time evolution of the general one-instanton
 solutions of our version of the Skyrme model in (2+1)-D,
 confirming some of the properties previously seen in other
 versions of the model. The extra terms added to the ordinary
 non-linear $O(3)$ model break its conformal invariance and
 stabilize the solitons. The changes of soliton size and position
 are negligible as time progresses.  Also, the time evolution
of the two-skyrmion configuration has showed us that, like in the
 previously studied model, the forces between the skyrmions
are repulsive.

 A paper on the scattering properties of our skyrmions will be
 available in not too distant a future.
\vspace{5 mm}

\large{\bf Acknowledgements}
\vspace{3 mm}

\normalsize
R. J. Cova is deeply indebted to
 {\em Universidad del Zulia \/{\em and}
 Fundaci\'{o}n Gran Mariscal de Ayacucho} for their joint
 grant supporting his {\em Ph.D.} studies  in Durham. He also
wishes to thank David Bull for helpful discussions.

\end{document}